\documentclass[10pt]{article}
\usepackage{algorithm}
\usepackage{subcaption}
\usepackage{multirow}
\usepackage[utf8]{inputenc}
\usepackage[T1]{fontenc}
\usepackage[margin=2.6cm]{geometry}
\usepackage{amsmath}
\usepackage{amssymb}
\usepackage{tcolorbox}
\usepackage{libertine}
\usepackage{tabto}
\usepackage{tikz}

\makeatletter
\newenvironment{breakablealgorithm}
  {
   \begin{center}
     \refstepcounter{algorithm}
     \hrule height.8pt depth0pt \kern2pt
     \renewcommand{\caption}[2][\relax]{
       {\raggedright\textbf{\fname@algorithm~\thealgorithm} ##2\par}%
       \ifx\relax##1\relax 
         \addcontentsline{loa}{algorithm}{\protect\numberline{\thealgorithm}##2}%
       \else 
         \addcontentsline{loa}{algorithm}{\protect\numberline{\thealgorithm}##1}%
       \fi
       \kern2pt\hrule\kern2pt
     }
  }{
     \kern2pt\hrule\relax
   \end{center}
  }
\makeatother

\begin{document}

\title{Exploring Behaviors of Hybrid Systems via the Voronoi Bias over Output Signals \thanks{This work was supported by the project GA21-09458S of the Czech Science Foundation GA \v{C}R and institutional support RVO:67985807.}}

\author{Gidon Ernst$^{1}$\footnote{ORCID: 0000-0002-3289-5764} \;and Ji{\v r}{\'\i} Fejlek$^{2}$\footnote{ORCID: 0000-0002-9498-3460} \\
$^1$ Software and Computational Systems Lab, Department of Computer Science, \\ Ludwig-Maximilians-Universität München, Oettingenstraße 67, Munich, 80538, Germany\\
$^2$The Czech Academy of Sciences, Institute of Computer Science,\\ Pod Vod\'{a}renskou v\v{e}\v{z}\'{i} 271/2, Prague, 182 07, Czech Republic}
\date{}

\maketitle

\begin{abstract}
In this paper, we consider an analysis of temporal properties of hybrid systems based on simulations, so-called falsification of requirements. We present a novel exploration-based algorithm for falsification of black-box models of hybrid systems based on the Voronoi bias in the output space. This approach is inspired by techniques used originally in motion planning: rapidly exploring random trees. Instead of commonly employed exploration that is based on coverage of inputs, the proposed algorithm aims to  cover all possible outputs directly. Compared to other state-of-the-art falsification tools, it also does not require robustness or other guidance metrics tied to a specific behavior that is being falsified. This allows our algorithm to falsify specifications for which robustness is not conclusive enough to guide the falsification procedure.
\end{abstract}

\section{Introduction}
In this paper, we consider the problem of testing temporal properties of black-box models of hybrid systems. The need to test properties of complex autonomous systems such as cyber-physical systems to detect possible errors and failures has been recognized by industry and academia worldwide~\cite{Corso:21}. Due to the complexity of these systems, formal verification methods are often not applicable, and thus, numerous algorithms and software tools for the analysis of black-box models based on simulations have been developed. These tools are compared in software competitions such as annually organized friendly competition for systems verification ARCH-COMP~\cite{arch}\footnote{https://cps-vo.org/group/ARCH/FriendlyCompetition}. In this context, testing properties and finding their violations is called \emph{falsification} of requirements. These requirements are described using temporal logic such as signal temporal logic (STL)~\cite{stl}. 

The most prominent approach in the state-of-the-art falsification tools is based on optimization~\cite{breach,psystaliro,athena,foresee} of robustness~\cite{stl}, a guidance metric associated with a given error STL specification that is being tested. However, a typical problem is that robustness is not conclusive enough to guide the optimization algorithm because robustness contains many local minima and flat regions~\cite{pant2017smooth} (e.g., due to the so-called scale problem~\cite{foresee}).

Since robustness may not provide any meaningful information for large regions of the input space, it is crucial for any falsification algorithm to explore the possible behaviors of the system efficiently. Exploration in the input space has been proposed in the past~\cite{adimoolam2017classification,crossentropy,falstar}. Input space exploration proceeds either by luck using random sampling or, exhaustively, by coverage. However, coverage in the input space does not necessarily coincide with coverage of the actual possible behaviors of the system, i.e., coverage in the output space. Not to mention the fact that reaching a sufficient level of coverage may require a prohibitively large number of samples. 

In this paper, we investigate an alternative way of exploration: exploration in the output space. Namely, we take inspiration from the methods used originally in robotics for motion planning: rapidly exploring random trees (RRT)~\cite{RRT1}. The RRT algorithm is able to solve difficult high-dimensional motion planning problems in complex environments with obstacles thanks to its ability to effectively explore the output space using so-called the Voronoi bias~\cite{RRT1}, which effectively steers the RRT algorithm to the unexplored parts of the output space. 
RRT algorithms have been proposed for falsification problems in the past~\cite{branicky2006sampling,dreossi2015efficient}, but their application in black-box model falsification was limited. RRTs require incremental simulations from various initial states of the black-box model to explore the whole output space, which often makes them computationally inefficient~\cite{dreossi2015efficient}. In addition, RRTs are able to detect consistently merely particular forms of the error behavior as we demonstrate in Section~\ref{sec:RRT_fals}.

\textbf{Contribution:} we propose a novel exploration-driven algorithm called \emph{Output Space Exploration} (OSE). It is inspired by the RRT algorithm but will alleviate the drawbacks of RRTs in the context of falsification and turn output space exploration into an effective tool for falsification. We demonstrate by extensive computational experiments that the resulting approach is not only competitive in comparison to other state-of-the-art falsifications methods but also complements them by solving falsifications problems that are challenging for the previous methods.

\section{Falsification: optimization and exploration}
Let us first formulate formally the falsification problem. We assume a deterministic black-box model that comprises continuous dynamics with discontinuities. We describe the model using an input/output function $\mathcal{M}\colon \mathcal{U} \mapsto \mathcal{Y}$, where $u \in \mathcal{U}$ is a real-valued piecewise continuous input signal $u\colon [0, T] \mapsto U$  of a length $T$ bounded to a compact set $U\subset \mathbb{R}^n$ and $y \in \mathcal{Y}$ is a real-valued piecewise continuous output signal $y\colon [0, T] \mapsto Y$ of the same length $T$ bounded to a compact set $Y\subset \mathbb{R}^m$. We further assume that the system specifications are described by the syntax of signal temporal logic (STL)~\cite{stl}
\begin{equation}
\phi ::= \neg \phi \mid \phi \vee \phi \mid \phi \wedge \phi   \mid \phi \mid \Box_J \phi \mid \Diamond_J \phi \mid \mu, 
\end{equation} 
where $\Box$ denotes the until temporal operator, $\Diamond$ denotes the eventually temporal operator, $J$ is the interval of time over which the temporal operators range, and $\mu$ are predicates over the dimensions $y_1, \ldots, y_m$ of an output signal $y$.

Assume a model $\mathcal{M}$, a time-bound $T$, an input bound $U$, and an STL specification $\phi$. In this paper, we consider the problem of finding a \emph{counterexample}, an input signal $u\colon [0, T] \mapsto U$ that causes the model $\mathcal{M}$ to violate the specification $\phi$, i.e., $\mathcal{M}(u)  \not\models\phi$. We will refer to this problem further as the \emph{falsification} problem. The most common approach for falsification of STL requirements of black-box models is based on optimization~\cite{breach,psystaliro,athena,foresee}, which detects design errors by minimizing the value of robustness~\cite{stl}. However, the robustness is, in general, difficult to optimize due to being often nonsmooth with many local minima and flat regions~\cite{pant2017smooth}. In such a case, the ability of optimization algorithms to find the counterexamples successfully is seriously deteriorated. 

To illustrate this point, let us assume an optimization algorithm CMA-ES~\cite{cmaes}, a predominantly local derivative-free population-based optimization method widely used in falsification tools~\cite{breach,falstar,foresee}. We consider the automatic transmission (AT) ARCH-COMP benchmark~\cite{arch} (see Section~\ref{sec:comp}) with an STL specification
$$ \Box_{[0, 29]} ((\mathrm{RPM} < 4770) \vee \Box_{[0, 1]} (\mathrm{RPM} >1000)).$$ 
As was observed in~\cite{foresee}, this specification is challenging for the CMA-ES algorithm. We initialized the CMA-ES optimization algorithm from 50 uniformly randomly selected inputs and merely 12 runs found the falsifying counterexample after 2500 model simulations. If we consider a modified STL formula from~\cite{foresee} 
\begin{multline*}
(\Box_{[0, 29]} (\mathrm{speed} < 100) \vee \Box_{[29, 30]} (\mathrm{speed} > 64))  \wedge \\ 
\Box_{[0, 29]} ((\mathrm{RPM} < 4770) \vee \Box_{[0, 1]} (\mathrm{RPM} >1000)),
\end{multline*}
the number of successful runs dropped to 0 due to the presence of the additional STL formula that adversarially changed the robustness even though the falsifying counterexample is the same. Another example of a specification in which robustness typically does not contain enough relevant information is the one that includes the discrete output signals, e.g.,
$$\neg ( \Diamond_{[0, 10]} (\mathrm{speed} > 50 ) \wedge \Diamond_{[10, 20]} (\mathrm{gear} < 2 ) \wedge \Diamond_{[20, 30]} (\mathrm{gear} > 3 ) ).$$
Due to the presence of a discrete output \emph{gear}, which takes only values from the set $\{1,2,3,4\}$, the robustness is constant on large regions of the input space. Hence, CMA-ES almost always immediately terminates due to observing no change in the value of robustness for the neighboring samples.

As demonstrated by these examples, robustness may not always provide good information to guide the falsification procedure. Thus, it is crucial for any falsification tool to explore the possible behaviors of the system independently of the robustness. A common way of exploration in the falsifying algorithms is input-based, using either purely randomized sampling  (e.g., uniform random sampling or nonuniform sampling in FalStar~\cite{falstar} ) or coverage-driven (e.g., Latin hypercube design~\cite{arch}). However, all of these input-based approaches may fail if the falsifying input signal is rare or even not included in the particular input-based search strategy.

Since exploration based merely on coverage of inputs can be unreliable, we propose an algorithm that aims to explore the output space directly and generate varied output signals that either falsify STL specifications by themselves or can serve as initializations for the optimization algorithms. To achieve that, we take an inspiration from the rapidly exploring random tree algorithm (RRT)~\cite{RRT1}, which was designed to cover all reachable outputs of dynamical systems to solve motion planning problems.

\section{Rapidly Exploring Random Trees}

In this paper, we introduce an output space exploration algorithm inspired by approaches that were originally used in robotics motion planning. In particular, we consider rapidly exploring random trees (RRT)~\cite{RRT1}. RRTs became successful at motion planning in robotics applications, finding solutions to high-dimensional motion planning problems in complex environments with obstacles~\cite{noreen2016optimal} for which other methods, such as grid-based methods or global optimization methods, would be too computationally expensive. In this section, we describe the original RRT algorithm and later demonstrate why the RRT algorithm is not suited for falsification problems in general.

\subsection{RRT algorithm}
\label{sec:RRT_fals}

\begin{figure}
\centering
\includegraphics[scale = 1.25]{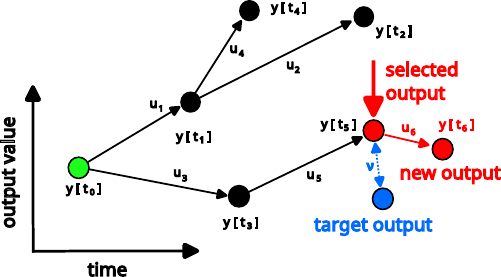}
\caption{Rapidly exploring random tree (Algorithm 1)}
\label{figure1}
\end{figure}

The RRT algorithm is able to solve these difficult motion planning problems thanks to its ability to explore the output space $Y \times [0, T]$. The RRT algorithm builds a random tree in the output space that gradually fills the output space, i.e., the tree nodes correspond to individual values of output signals. The tree is built incrementally from an initial output, and its growth is driven to the unvisited parts of the output space.

Assume a black-box model $\mathcal{M}$ and let the initial value of the output signal be $y(t_0)$ at the initial time $t_0 = 0$. We denote a pair $(y(t_0),t_0)$ in the output space $Y \times [0, T]$ as $y[t_0]$. In each step, the RRT algorithm further grows the tree by concatenating some previous input signals with new randomly generated input signals.

The key aspect of the RRT algorithm is the selection of input signals for further extension, which is done by randomly generating an output value and picking the node in the tree that is closest to it with respect to some appropriately chosen metric $\nu$, e.g., the Euclidean metric, see Figure~\ref{figure1}. We will refer to this hypothetical output further as \emph{target} output. Thanks to this selection procedure, the probability of choosing any given output in the tree for further expansion is proportional to the volume of a set of target outputs that are the closest to this output. Thus, it is more likely to select outputs that are more isolated, i.e., located in a more unexplored area of the output space. This phenomenon is called the \emph{Voronoi bias}~\cite{RRT2}.

Algorithm 1 summarizes the whole RRT algorithm. We assume that the targets are generated from the output space $Y \times [0,T]$, which we wish to explore. The inputs are generated from the set of real-valued piecewise continuous input signals  $\mathcal{U}$. For example, inputs can be generated as constant signals of a given value and length, i.e., the resulting concatenated input signals are piecewise constant. Overall, the parameters of the algorithm are as follows

\smallskip
\begin{itemize}
\item A distribution $p_Y$ on $Y \times [0,T]$ for generating random target points in the Euclidean plane
\item A distribution $p_\mathcal{U}$ on $\mathcal{U}$ for generating random inputs 
\item A metric $\nu$ on $Y\times [0,T]$
\end{itemize}
\smallskip

The resulting RRT algorithm has the following form~\cite{RRT1}.
\smallskip
\begin{breakablealgorithm}
\caption{Rapidly Exploring Random Tree Algorithm}
\begin{itemize}
\item [IN:] A black-box model $\mathcal{M}$, a~bounded set of input values $U$, a~bounded time horizon $T$
\item [OUT:] A tree $\mathcal{T}$ in the output space $Y\times [0,T]$
\end{itemize}

\begin{flushleft}
Repeat:
\end{flushleft}

\begin{itemize}
\item[(a)] Generate a target output $(y',t')$ from $p_Y$ (\emph{target selection})
\item[(b)] Select a node $y[t]$ from $\mathcal{T}$ that is the closest to $(y',t')$ wrt~$\nu$  (\emph{output selection})
\item[(c)] Generate a random input $u'$ from $p_\mathcal{U}$ (\emph{input selection})
\item[(d)] Add a new node $y_{\mathrm{new}} =  \mathcal{M}(uu')$ to $\mathcal{T}$, where $y = \mathcal{M}(u)$ (\emph{simulation}) 
\end{itemize}
\end{breakablealgorithm}
\smallskip
We identify three key properties that make this algorithm promising for falsification
\begin{enumerate}
\item[(1)] the RRT algorithm merely uses system simulations to expand the tree of reachable outputs
\item[(2)] the RRT algorithm does not need any cost/objective function to steer exploration
\item[(3)] the RRT algorithm explores the output space $Y \times [0,T]$ 
\end{enumerate} 
Property (1) makes Algorithm 1 suitable for use on a black-box model since the RRT algorithm does not require some deeper knowledge of its inner dynamics apart from general knowledge of reachable values of output signals to determine $Y$. Property (2) enables running the RRT algorithm without any reliance on the robustness or other STL-based guidance metrics in searching for falsifying counterexamples, which is otherwise a necessity when using optimization-based approaches for falsification. In addition, it enables testing multiple STL specifications simultaneously. However, Property (3) is insufficient for falsification in general due to the restriction of exploration of $Y \times [0,T]$, as we demonstrate in the following subsection.

\subsection{RRT and model falsification}

RRTs have been proposed for falsification problems in the past~\cite{branicky2006sampling,dreossi2015efficient}, but there are challenges encountered in their use for black-box model falsification. RRTs require incremental simulations from various initial states of the black-box model to explore the whole output space. However, a black-box model cannot be, in general, simulated from an arbitrarily chosen initial state apart from the one from which it is meant to be initialized. Hence, to continue where some previous simulation ended, the algorithm either has to repeat the whole simulation as a whole or the algorithm has to have access to a "suspended" ongoing simulation. This second option is often very impractical to implement as observed in~\cite{falstar}. Concerning the first option, the average length of a partial simulation is half of the time horizon $T$ provided that time instants of targets are sampled uniformly. Thus, the computational time of simulations is approximately halved, and in some cases, it can be even less, e.g., due to the compilation step before each simulation~\cite{dreossi2015efficient}.

To conclude, the difference between a partial and a full simulation can be actually quite small in practice. This observation, together with a fact, that a new incremental simulation usually does not contain as much of new information as a brand new full-length one would (provided that the incremental step is comparatively small like in~\cite{branicky2006sampling,dreossi2015efficient}), causes the RRT algorithm to converge slowly and to be overall computationally inefficient for black-box falsification compared to the algorithms that operate only with full-length simulations, as we demonstrate further in our computational experiments in Section 5. 

\begin{figure}
\centering
\includegraphics[scale = 0.6]{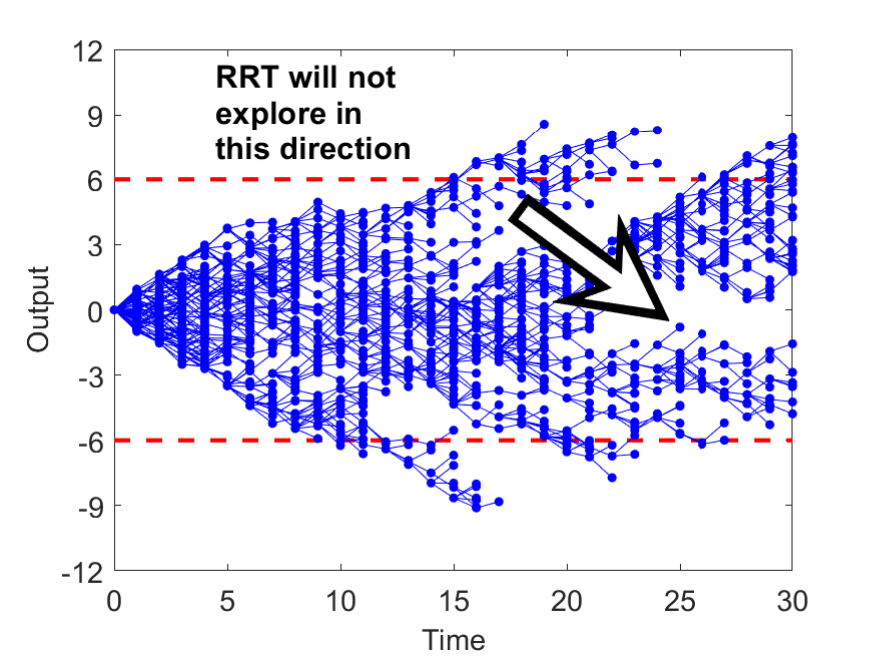}
\caption{Progress of Algorithm 1 for the simple integrator problem. The RRT algorithm will block itself from exploring further trajectories that cross those already seen, i.e., it will not be able to follow the indicated arrow.}
\label{figure2}
\end{figure}

In addition, RRTs are limited by the form of model behaviors that they can search effectively for. The output selection strategy in the RRT algorithm achieves output space exploration. But this exploration is reduced to generating all possibly reachable \emph{individual output values} $y[t]$ from $Y\times[0, T]$ but does not include exploration of \emph{all possible output signals} $y\colon [0, T] \mapsto Y$. While the output selection strategy in the RRT algorithm favors the exploration towards the unexplored regions of $Y\times[0, T]$, such a strategy also inherently limits further expansions of the tree in the already covered parts of $Y\times[0, T]$~\cite{RRT5}. 

To illustrate this phenomenon, let us assume a simple integrator problem: the model $\mathcal{M}$ is governed by the ordinary differential equation $\dot{y} = u$, where $u \in [-1, 1]$, and let $T = 30$ and $y(0) = 0$. We pick $Y = [-10, 10]$, $p_Y$ as the uniform distribution on $[-10, 10] \times [0, 30]$ and we pick a fixed-time step input selection, i.e., $p_u$ is the uniform distribution on $U = [-1, 1]$ but the length  of the input signal is constant $h = 1$. Concerning metric $\nu$, we use the Euclidean metric on $[-10, 10] \times [0, 30]$. The progress of Algorithm 1 for this example can be seen in Figure~\ref{figure2}. Algorithm 1 starts to fill out the parts of $[-10, 10] \times [0, 30]$ that are reachable for $\mathcal{M}$. However, Algorithm 1 does not systematically explore other ways of reaching particular outputs apart from the ones discovered earlier. Instead, Algorithm 1 mostly only fills out the gaps. 

This behavior has serious drawbacks when simulating output signals for falsification as indicated in Figure~\ref{figure2}. For example, let us consider an STL specification $\phi = \left(\Diamond_{[0, 30]}(y > 6) \Rightarrow \Box_{[0, 30]}(y>-6)\right)$. Falsification of $\phi$ requires an output signal that initially rises to value $6$ and subsequently goes below $-6$ (or the other way around). Unless Algorithm 1 generates the tree with this shape of the output signal early on, it will be unable to do so later because the earlier parts of the tree will block further expansions that would go "through" these earlier parts of the tree. Hence, it is very unlikely that Algorithm 1 will find a falsifying counterexample to the STL formula $\phi$ in any given run. 

Consequently, we propose an innovative exploration algorithm based on the original RRT algorithm that will alleviate both crucial drawbacks of RRTs and that be an effective algorithm for falsification. We achieve this by expanding the RRTs from the output space $Y\times [0, T]$ into the \emph{space of output signals} $\mathcal{Y}$.

\section{Exploration in the space of output signals} 

In this section, we propose a novel exploration-based algorithm for falsification to complement optimization-based falsification, especially when the robustness metric is not conclusive enough to guide the optimization algorithm to the falsifying output. In addition, since the exploration will be done in the output space and not in the input space, it also complements falsification strategies based on a search in the input space.

Our construction of this algorithm is inspired by the original RRT algorithm (Algorithm 1).  As we discussed in the previous section, the RRT algorithm explores the output space $Y\times [0, T]$ by incrementally simulating the model $\mathcal{M}$ and building a tree of reachable outputs from $Y\times [0, T]$. However, such a method of exploration may not be enough to falsify STL requirements that encompass more varied behavior than reaching a particular output value. To address the challenges of using the RRT algorithm for the falsification that are encountered due to the transition from motion planning, we suggest moving the exploration from the output space $Y\times [0, T]$ to the space of the output signals $\mathcal{Y} = \{y\colon [0, T] \mapsto Y \mid y \text{ is piecewise continuous}\}$. This allows this new algorithm to keep exploring even once the reachable parts $Y\times [0, T]$ are covered. 

In addition, since this new algorithm operates in the space of the output signals $\mathcal{Y}$, it no longer employs incremental simulations of the model. This change alone dramatically increases the performance for falsification, as we demonstrate in our computational experiments in Section 5.

\subsection{Conceptual Algorithm}

To build an exploration algorithm in the space of output signals $\mathcal{Y}$ inspired by the RRT algorithm, we must redefine the target and output selection steps. Generating random signals from $\mathcal{Y}$ that are somewhat similar to the actual output signals $y$ from the model $\mathcal{M}$ to serve as targets that steer the exploration algorithm is nontrivial without some deeper knowledge of the inner dynamics of $\mathcal{M}$. Consequently, we will generate targets from a far simpler set than $\mathcal{Y}$. 

Formally, we first select a \emph{feature space} $\mathcal{F}$ and a distribution on $p_\mathcal{F}$ that generates random target features. Concerning the output selection, we assume a mapping $f\colon\mathcal{Y} \times \mathcal{F} \mapsto [0, +\infty]$ that evaluates the distance between a signal $y$ and a randomly generated target feature. We will refer to $f$ further as the \emph{feature metric}.

In addition, since inputs are no longer incrementally concatenated, the distribution for generating new inputs must also generate whole random input signals  $u'\colon [0, T] \mapsto U$. We will use the input selection rule 
\begin{equation}
u' = u + \delta, 
\end{equation}
where $u$ is an input corresponding to the selected output and $\delta\in \mathcal{U}$ is a random input disturbance generated from a chosen distribution $p_\mathcal{U}$. This modification provides a greater variance in the generation of new inputs in comparison to the original RRT algorithm, in which the input signal stays the same apart from random input that is concatenated at the end. As we demonstrate in the computational experiments, this dramatically aids the resulting algorithm to generate a diverse set of output signals quickly.

Algorithm 2 summarizes the conceptual exploration algorithm in the space of output signals. The parameters of the algorithm are as follows

\smallskip
\begin{itemize}
\item A feature space $\mathcal{F}$, a feature metric $f\colon\mathcal{Y} \times \mathcal{F} \mapsto [0, +\infty]$
\item A distribution $p_\mathcal{F}$ on $\mathcal{F}$ for generating random target features
\item A distribution $p_\mathcal{U}$ for generating random input disturbances 
\end{itemize}
\smallskip

\begin{breakablealgorithm}
\caption{Conceptual Exploration Algorithm in the Space of Output Signals based on the Voronoi Bias}
\begin{itemize}
\item [IN:] A black-box model $\mathcal{M}$, a~bounded set of input values $U$, a~bounded time horizon $T$
\item [OUT:] A library of output signals $\mathcal{L}$
\end{itemize}
\begin{flushleft}
Repeat:
\end{flushleft}
\begin{itemize}
\item[(a)] Generate a target feature $F'$ from $p_\mathcal{F}$ (\emph{target selection})
\item[(b)] Select an output signal  $y = \mathcal{M}(u)$ from $\mathcal{L}$ that is closest to~$F'$ wrt~$f$ (\emph{output selection})
\item[(c)] Generate a random input disturbance $\delta$ from $p_\mathcal{U}$ and let $u' = u + \delta$   (\emph{input selection})
\item[(d)] Add a new node $y_{\mathrm{new}} =  \mathcal{M}(u')$ to $\mathcal{L}$ (\emph{simulation})
\end{itemize}
\end{breakablealgorithm}
\smallskip

\subsection{Implementation}

In this subsection, we describe our implementation of Algorithm~2. Let us first describe the feature space which our implementation uses. Let $y_1, \ldots, y_m$ be individual components of an output signal $y$, i.e., $y = (y_1, \ldots, y_m)$, where $y_i\colon [0, T] \mapsto Y_i$ and $Y = Y_1 \times Y_2 \times \cdots \times Y_m.$ We denote a set $Y_I = Y_{I_1}\times \cdots \times Y_{I_k}$, where $I$ is a set of indices $I = \{I_1, \ldots, I_k\} \subset \{1, \ldots, m\}$ such that $I_1 < \cdots < I_k$.

In our implementation, we instantiate Algorithm 2 using the feature spaces 
$$\mathcal{F}^I_k = \left\{(F_1^I, \ldots, F_k^I) \mid F_1^I, \ldots, F_k^I \in Y_I \times [0, T]\right\},$$
Thus, the feature of the $k$th \emph{level} is a set of points $F^I_1, \ldots, F^I_k$ from $Y_I \times [0, T]$. The set of indices $I$ corresponds to the output components, according to which is the exploration performed.

The level $k = 1$ with the indices $I$ corresponds to the features given by a single value of the output from  $Y^I \times [0,T]$. Hence, Algorithm 2 with the feature space $\mathcal{F}^I_1$ is a direct analog to Algorithm 1, but with the new input selection step and without the incremental simulations. In our implementation, we consider additional levels $k > 1$ that increase the complexity of the exploration strategy by distinguishing output signals that approximately visit multiple output values in multiple time instants. 

\paragraph{Selection of feature levels.}  Our main implementation uses the feature spaces  $\mathcal{F}^I_k$ for $k = 1,2,3.$ The components $I$ are selected for each particular benchmark. In each iteration, the feature space $\mathcal{F}_k$ is chosen randomly with probability $p_1 = \frac{4}{7}, p_2 = \frac{2}{7}$, and $p_3 = \frac{1}{7}$. We choose this heuristic to achieve greater consistency in selecting extreme-valued output signals which emphasize exploration outward, outside the set of already visited outputs. We investigate this choice further in the first part of our computational experiments in Section 5.

\paragraph{Generation of random features.}  We generate random features from an estimate of the set of output signal values. Namely, we generate a random target feature $F^{\{1,\ldots,m\}}$ of level $k = 1$ at a time instant $t$ as
\begin{equation}
\label{est}
F^{\{1,\ldots,m\}} = 1.4\left(\mathrm{UB}(t) - \mathrm{LB}(t)\right)\mathrm{rand} + \mathrm{LB}(t) - 0.2\left(\mathrm{UB}(t) - \mathrm{LB}(t)\right),
\end{equation}
where $\mathrm{rand}$ is uniformly distributed on $[0,1]^m$ and $\mathrm{UB}(t)$ and $\mathrm{LB}(t)$ are the maximum and the minimum observed output signal values at the time instant $t$ respectively. The time instant $t$ is generated uniformly randomly from $[0, T]$. Thus, random features at the time instant $t$ are generated from a slightly larger set $[\mathrm{LB}(t) - 1.2(\mathrm{UB}(t) - \mathrm{LB}(t)), \mathrm{UB}(t) + 1.2(\mathrm{UB}(t) - \mathrm{LB}(t))]$ than the observed possible values of the output signal $[\mathrm{LB}(t), \mathrm{UB}(t)]$. We choose this heuristic to emphasize the exploration outward from the set $[\mathrm{LB}(t), \mathrm{UB}(t)]$, since features outside of $[\mathrm{LB}(t), \mathrm{UB}(t)]$ bias the selection of outputs in favor of the extreme-valued ones.

\paragraph{Feature metric.} Concerning the feature metric on the feature spaces $\mathcal{F}_k$, we picked the feature metric 
\begin{equation}
f_k(y,F^I) = \sum_{j = 1}^k\sum_{i \in I} \left(\frac{y_i(t_j) - y_i^j}{\mathrm{UB}_i(t_j) - \mathrm{LB}_i(t_j)}\right)^2
\end{equation}
where $F^I = (F^I_1, \ldots, F^I_k)$ and $F^I_i = (y^i, t_i)$ for all $i = 1 \ldots, k$. Hence, we compute a weighted Euclidean distance between a random feature $(y',t)$ and a signal output value $y(t)$ for the corresponding time instant $t$ in dimensions $I$ and add them all together. We scale the difference wrt the maximum observed difference in a given dimension of the output signal in the given time instant to alleviate adding up values of vastly different scales.

\paragraph{Generation of random inputs.}  We assume that investigated inputs $u$ are piecewise constant with a fixed time step $h$, which provides a straightforward finite dimensional parametrization of random disturbances $\delta$.  Let us assume that the input signals are given by values $u_1, \ldots u_l \in U$. We use an approach that is not dissimilar to global optimization algorithms such as genetic algorithms or differential evolution~\cite{de}. For simplicity's sake, let us assume one-dimensional input signals. We select a \emph{crossover probability} $0 < CR < 1$ and for each $i = 1, \ldots, l $ we modify the $i$th element of the output with the probability $CR$ 
\begin{equation}
\label{mod}
u'_i = u_i + \delta_i, 
\end{equation}
where $\delta = \gamma \tan \left( \pi (\mathrm{rand} - 0.5)\right)$, where $\mathrm{rand}$ is uniformly distributed from $[-1,1]$, i.e., the disturbance $\delta$ follows the Cauchy distribution with the scale parameter $\gamma > 0$. For a multidimensional input signal, we use the same input selection and apply the disturbance in \eqref{mod} per dimension. We select the scale parameter $\gamma$ as a quarter of the overall range of the admissible input values and the crossover probability $CR = 0.5$.

Algorithm 3 (Output Space Exploration, OSE) summarizes the implementation of Algorithm 2. To demonstrate our implementation, let us return to the simple integrator problem. Since the model has one-dimensional output, we pick $I = \{1\}.$ As can be seen in Figure 3, OSE continues to explore in comparison to RRTs (Algorithm 1) and eventually finds the falsifying trajectory.

\begin{algorithm}
\caption{Output Space Exploration (OSE)}
\begin{itemize}
\item [IN:] A black-box model $\mathcal{M}$, a~bounded set of input values $U$, a~bounded time horizon $T$, indices $I$ of components of the output space according to which is the exploration performed
\item [OUT:] A library of output signals $\mathcal{L}$
\end{itemize}
\begin{flushleft}
Repeat
\end{flushleft}
\begin{itemize}
\item[(a)] Randomly pick a level of the features with probabilities $p_1 = \frac{4}{7}, p_2 = \frac{2}{7}$, and $p_3 = \frac{1}{7}$. Let $k$ be a chosen level, generate $k$ random target features $F^I_1, \ldots F^I_k$ from $Y_I \times [0,T]$ using  distribution \eqref{est} (\emph{target selection})
\item[(b)] Select an output signal $y = \mathcal{M}(u)$ from $\mathcal{L}$ that is closest to $F^I = \{F^I_1, \ldots F^I_k\}$ according to the feature metric
$$ f_k(y,F^I) = \sum_{j = 1}^k\sum_{i \in I} \left(\frac{y_i(t_j) - y_i^j}{\mathrm{UB}_i(t_j) - \mathrm{LB}_i(t_j)}\right)^2,$$
where $F^I_j = (y^j, t_j)$ for all $j = 1 \ldots, k$ (\emph{output selection})
\item[(c)] Generate a random input disturbance $\delta$ according to \eqref{mod} and let $u' = u + \delta$ (\emph{input selection})
\item[(d)] Add a new output signal $y_{\mathrm{new}} =  \mathcal{M}(u')$ to $\mathcal{L}$ (\emph{simulation})
\end{itemize}
\end{algorithm}

\begin{figure}
\centering
\includegraphics[scale = 0.65]{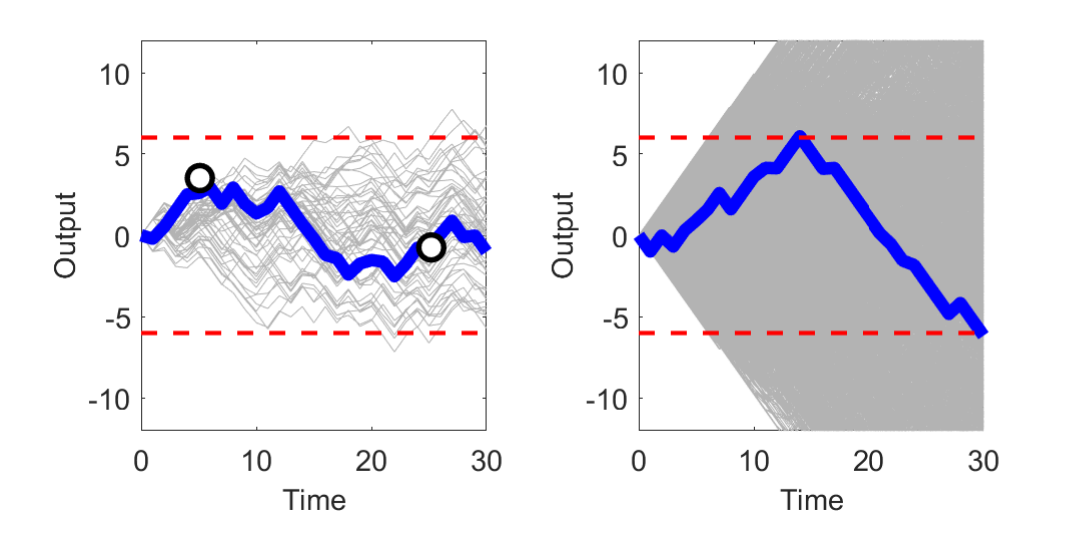}
\caption{Progress of OSE (Algorithm 3) for the simple integrator. Black-bordered points denote a feature by which the current output signal (blue) was selected. The second figure shows the falsifying output signal found by OSE.}
\label{figure3}
\end{figure}

\section{Computational experiments}
\label{sec:comp}

In this section, we describe our computational experiments and provide results that demonstrate the capabilities of OSE (Algorithm 3). We will demonstrate its exploration capabilities in comparison to other exploration-based algorithms. We will also demonstrate its performance in comparison to the optimization-based approaches.

The whole implementation was done in MATLAB 2020b. The computational experiments were performed on Intel Core i5-11320H @ 3.20GHz 4-core processor with 16GB RAM.

\subsection{Benchmarks}

First, we list all falsification benchmark problems used in the computational experiments. The outputs from all black-box models were re-sampled using linear interpolation in such a way to produce output signals with a fixed time step $0.1$. The STL specifications for the resulting output signals were evaluated discretely using the same fixed time step $0.1$.

\paragraph{Simple Integrator} We assume  a model that is governed by the ordinary differential equation $\dot{y} = u$ with input specification $-1 \leq u \leq 1$.  Input signals are assumed to be piecewise constant with discontinuities at most every 30 time units.

{\small
\allowdisplaybreaks
\begin{align*}
\text{SI1}:\; & \Box_{[0, 30]} ( y < 25 ) \\
\text{SI2}:\; & \Box_{[0, 30]} ( y > -25 ) \\
\text{SI3}:\; & \Box_{[0, 15]} ( y < 12.5 ) \\
\text{SI4}:\; & \Box_{[0, 15]} ( y < -12.5 )\\
\text{SI5}:\; & \Diamond_{[0, 30]} ( |y| > -1.3 )\\
\text{SI6}:\; & (\Diamond_{[0, 10]}( y < -6)) \Rightarrow (\Diamond_{[25, 30]} ( |y+5|>1.3))\\
\text{SI7}:\; & \Diamond_{[10, 30]} ( |y+5| > -1.3 ) \\
\text{SI8}:\; & \Diamond_{[20, 30]} ( |y-15| > -1.3 ) \\
\text{SI9}:\; & \Diamond_{[20, 30]} ( |y+15| > -1.3 ) \\
\text{SI10}:\; & (\Diamond_{[0, 30]} ( y > 6)) \Rightarrow (\Box_{[0, 30]} ( y>-6)) \\
\text{SI11}:\; & (\Box_{[0, 15]} ( |y| < 1.5)) \Rightarrow (\Box_{[0, 30]} ( |y| < 10 )) \\
\text{SI12}:\; & (\Box_{[0, 10]}(|y| < 2.5) \wedge \Box_{[20, 30]}( |y| < 2.5)) \Rightarrow (\Box_{[10, 20]} ( |y| < 5)) \\
\text{SI13}:\; & (\Diamond_{[0, 10]}( |y| > 8 )  \wedge \Diamond_{[20, 30]}(|y| > 8)) \Rightarrow (\Box_{[10, 20]} ( |y| < 4))\\
\text{SI14}:\; & (\Box_{[10, 20]}( y > 6 )  \wedge \Box_{[10, 20]}(y < 9))   \Rightarrow (\Box_{[20, 30]} ( y < 14))\\
\text{SI15}:\; & (\Diamond_{[0, 10]} ( y > 6)) \Rightarrow (\Diamond_{[25, 30]} ( |y-5|>1.3))\\
\text{SI16}:\; & (\Diamond_{[0, 10]} ( y < -6)) \Rightarrow (\Diamond_{[25, 30]} ( |y+5|>1.3)) 
\end{align*}
}

\paragraph{Automatic Transmisssion} This is a model of an automatic transmission and encompasses a controller that selects the gear 1 to 4 depending on two inputs (throttle, brake) and the current engine load, rotations per minute, and car speed. It is a standard falsification benchmark used in the competition ARCH-COMP. Input specification: $0 \leq \mathrm{throttle} \leq 100$ and $0 \leq \mathrm{brake} \leq 325$. Input signals are assumed to be piecewise constant with discontinuities at most every 5 time units. The STL specifications for falsification are chosen as follows. Specifications AT1, AT15--AT18, AT24, AT25--AT27 are taken directly from ARCH-COMP, and other specifications are inspired by~\cite {foresee}.

{\small
\allowdisplaybreaks
\begin{align*}
\text{AT1}:\; & \Box_{[0, 20]} ( \mathrm{speed} < 120 )\\
\text{AT2}:\; & \Box_{[0, 30]} ( \mathrm{speed} < 135 )\\
\text{AT3}:\; & \Diamond_{[0, 30]} ( \mathrm{speed} > 5 )\\
\text{AT4}:\; & \Diamond_{[15, 30]} (( \mathrm{speed} < 10 ) \vee ( \mathrm{speed} > 20 ))\\
\text{AT5}:\; & \Diamond_{[15, 30]} (( \mathrm{speed} < 30 ) \vee ( \mathrm{speed} > 40 ))\\
\text{AT6}:\; & \Diamond_{[15, 30]} (( \mathrm{speed} < 50 ) \vee ( \mathrm{speed} > 60 ))\\
\text{AT7}:\; & \Diamond_{[15, 30]} (( \mathrm{speed} < 70 ) \vee ( \mathrm{speed} > 80 ))\\
\text{AT8}:\; & \Diamond_{[15, 30]} (( \mathrm{speed} < 90 ) \vee ( \mathrm{speed} > 100 ))\\
\text{AT9}:\;  & \Diamond_{[0, 15]} ( \mathrm{speed} > 80 ) \Rightarrow \Box_{[15, 30]} ( \mathrm{speed} > 45 )\\
\text{AT10}:\; & \Box_{[0, 15]} ( \mathrm{speed} < 20 ) \Rightarrow \Box_{[15, 30]} ( \mathrm{speed} < 100 )\\
\text{AT11}:\; & \Box_{[0, 30]} ( (\mathrm{gear} = 4) \Rightarrow (\mathrm{speed} > 35 ) )\\
\text{AT12}:\; & \Box_{[0, 25]} ( (\mathrm{gear} = 4) \Rightarrow \Diamond_{[0, 5]} (\mathrm{RPM} < 4200 ) )\\
\text{AT13}:\; & \Box_{[0, 30]} ( (\mathrm{gear} = 2) \Rightarrow (\mathrm{speed} > 5 ) )\\
\text{AT14}:\; & \Box_{[0, 25]} ( (\mathrm{gear} = 2) \Rightarrow \Diamond_{[0, 5]} (\mathrm{RPM} < 3630 ) )\\
\text{AT15}:\; & \Box_{[0, 30]} (( (\mathrm{gear} \neq 1) \wedge \Diamond_{[0.01, 0.1]} (\mathrm{gear} = 1))\Rightarrow \Diamond_{[0.01, 0.1]}\Box_{[0, 2.5]} (\mathrm{gear} = 1) )) \\
\text{AT16}:\; & \Box_{[0, 30]} (( (\mathrm{gear} \neq 2) \wedge \Diamond_{[0.01, 0.1]} (\mathrm{gear} = 2))\Rightarrow \Diamond_{[0.01, 0.1]}\Box_{[0, 2.5]} (\mathrm{gear} = 2) )) \\
\text{AT17}:\; & \Box_{[0, 30]} (( (\mathrm{gear} \neq 3) \wedge \Diamond_{[0.01, 0.1]} (\mathrm{gear} = 3))\Rightarrow \Diamond_{[0.01, 0.1]}\Box_{[0, 2.5]} (\mathrm{gear} = 3) )) \\
\text{AT18}:\; & \Box_{[0, 30]} (( (\mathrm{gear} \neq 4) \wedge \Diamond_{[0.01, 0.1]} (\mathrm{gear} = 4))\Rightarrow \Diamond_{[0.01, 0.1]}\Box_{[0, 2.5]} (\mathrm{gear} = 4) )) \\
\text{AT19}:\; & \Diamond_{[0, 30]} ( \mathrm{gear} \neq 1 )\\
\text{AT20}:\; & \Diamond_{[15, 30]} ( \mathrm{gear} \neq 2 )\\
\text{AT21}:\; & \Diamond_{[15, 30]} ( \mathrm{gear} \neq 3 )\\
\text{AT22}:\; & \Diamond_{[15, 30]} ( \mathrm{gear} \neq 4 )\\
\text{AT23}:\; & \Box_{[0, 25]} ( ( \mathrm{gear} = 4 ) \Rightarrow \Box_{[0, 5]} ( \mathrm{gear} \neq 1  ))\\
\text{AT24}:\; & \Box_{[0, 30]} ( \mathrm{RPM} < 4770 )\\
\text{AT25}:\; & (\Box_{[0, 30]} (\mathrm{RPM} < 3000)) \Rightarrow (\Box_{[0, 4]} (\mathrm{speed} < 35 )) \\
\text{AT26}:\; & (\Box_{[0, 30]} (\mathrm{RPM} < 3000)) \Rightarrow (\Box_{[0, 8]} (\mathrm{speed} < 50 )) \\
\text{AT27}:\; & (\Box_{[0, 30]} (\mathrm{RPM} < 3000)) \Rightarrow (\Box_{[0, 20]} (\mathrm{speed} < 65 )) \\
\text{AT28}:\; & (\Box_{[0, 30]} (\mathrm{RPM} < 2700)) \Rightarrow (\Box_{[0, 10]} (\mathrm{speed} < 50 )) \\
\text{AT29}:\; & (\Box_{[0, 30]} (\mathrm{RPM} < 4500)) \Rightarrow (\Box_{[0, 10]} (\mathrm{speed} < 80 )) \\
\text{AT30}:\; & (\Box_{[0, 20]} ((\mathrm{RPM} < 650) \Rightarrow  \Diamond_{[0, 10]} (\mathrm{gear} \neq 1) ) \\
\text{AT31}:\; & \Box_{[0, 20]} (\Diamond_{[0, 10]} ((\mathrm{RPM} >3500) \vee (\mathrm{speed} < 115) ) )\\
\text{AT32}:\; & \Box_{[0, 29]} ((\mathrm{RPM} < 4770) \vee \Box_{[0, 1]} (\mathrm{RPM} >1000))\\
\text{AT33}:\; & \neg ( \Diamond_{[0, 10]} (\mathrm{speed} > 30 ) \wedge \Diamond_{[10, 20]} (\mathrm{speed} < 10 ) \wedge \Diamond_{[20, 30]} (\mathrm{speed} > 70 ) )\\
\text{AT34}:\; & \neg ( \Diamond_{[0, 10]} (\mathrm{gear} > 3 ) \wedge \Diamond_{[10, 20]} (\mathrm{gear} < 2 ) \wedge \Diamond_{[20, 30]} (\mathrm{gear} > 3 ) )\\
\text{AT35}:\; & \neg ( \Diamond_{[0, 10]} (\mathrm{speed} > 50 ) \wedge \Diamond_{[10, 20]} (\mathrm{gear} < 2 ) \wedge \Diamond_{[20, 30]} (\mathrm{gear} > 3 ) )\\
\text{AT36}:\; & \neg ( \Box_{[5, 10]} (\mathrm{gear} = 2 ) \wedge \Box_{[15, 20]} (\mathrm{gear} = 3 ) \wedge \Box_{[25, 30]} (\mathrm{gear} = 4 ) )\\
\text{AT37}:\; & \neg ( \Box_{[5, 10]} (\mathrm{gear} = 2 ) \wedge \Box_{[15, 20]} (\mathrm{gear} = 4 ) \wedge \Box_{[25, 30]} (\mathrm{gear} = 3 ) )\\
\text{AT38}:\; & \neg ( \Box_{[5, 10]} (\mathrm{gear} = 3 ) \wedge \Box_{[15, 20]} (\mathrm{gear} = 2 ) \wedge \Box_{[25, 30]} (\mathrm{gear} = 1 ) )\\
\text{AT39}:\; & \neg ( \Box_{[5, 10]} (\mathrm{gear} = 3 ) \wedge \Box_{[15, 20]} (\mathrm{gear} = 2 ) \wedge \Box_{[25, 30]} (\mathrm{gear} = 3 ) )\\
\text{AT40}:\; & \neg ( \Box_{[5, 10]} (\mathrm{gear} = 2 ) \wedge \Box_{[15, 20]} (\mathrm{gear} = 1 ) \wedge \Box_{[25, 30]} (\mathrm{gear} = 4 ) )
\end{align*}
}

\paragraph{Chasing Cars}  The model is derived from~\cite{hu2000towards}, which presents a simple model of an automatic chasing car. The chasing cars model consists of five cars, in which the first car is driven by inputs (throttle and brake), and the other four are driven by the algorithm from~\cite{hu2000towards}. The output of the system is the location of five cars $y_1, y_2, y_3, y_4, y_5$.  It is a standard falsification benchmark used in the competition ARCH-COMP. Input specification: $0 \leq \mathrm{throttle} \leq 1$ and $0 \leq \mathrm{brake} \leq 1$. Input signals are assumed to be piecewise constant with discontinuities at most every 20 time units. The STL specifications are inspired by the specifications in ARCH-COMP.

{\small
\allowdisplaybreaks
\begin{align*}
\text{CC1}:\; & \Box_{[0.0, 100.0]} ( y_5 - y_4 < 120.0 )\\
\text{CC2}:\; & \Box_{[0.0, 100.0]} ( y_4 - y_3 < 120.0 )\\
\text{CC3}:\; & \Box_{[0.0, 100.0]} ( y_3 - y_2 < 120.0 )\\
\text{CC4}:\; & \Box_{[0.0, 100.0]} ( y_2 - y_1 < 120.0 )\\
\text{CC5}:\; & \Box_{[0.0, 70.0]}\Diamond_{[0.0, 30.0]}( y_5 - y_4 > 12.0 )\\
\text{CC6}:\; & \Box_{[0.0, 70.0]}\Diamond_{[0.0, 30.0]}( y_4 - y_3 > 12.0 )\\
\text{CC7}:\; & \Box_{[0.0, 70.0]}\Diamond_{[0.0, 30.0]}( y_3 - y_2 > 12.0 )\\
\text{CC8}:\; & \Box_{[0.0, 70.0]}\Diamond_{[0.0, 30.0]}( y_2 - y_1 > 12.0 )\\
\text{CC9}:\; & (\Box_{[0.0, 100.0]} ( y_5 - y_4 < 80.0 ))\vee(\Box_{[0.0, 70.0]}\Diamond_{[0.0, 30.0]}( y_2 - y_1 > 15.0 ))\\
\text{CC10}:\; & (\Box_{[0.0, 100.0]} ( y_2 - y_1 < 35.0 ))\vee(\Box_{[0.0, 70.0]}\Diamond_{[0.0, 30.0]}( y_5 - y_4 > 15.0 ))\\
\text{CC11}:\; & (\Box_{[0.0, 100.0]} ( y_4 - y_3 < 50.0 ))\vee(\Box_{[0.0, 70.0]}\Diamond_{[0.0, 30.0]}( y_3 - y_2 > 15.0 ))\\
\text{CC12}:\; & (\Box_{[0.0, 100.0]} ( y_3 - y_2 < 50.0 ))\vee(\Box_{[0.0, 70.0]}\Diamond_{[0.0, 30.0]}( y_4 - y_3 > 15.0 ))\\
\text{CC13}:\; & \Box_{[0.0, 80.0]} ((\Diamond_{[0.0, 20.0]}(y_2 - y_1 < 45.0) )  \vee   (\Diamond_{[0.0, 20.0]}(y_5 - y_4 > 20.0) )\\
\text{CC14}:\; & \Box_{[0.0, 80.0]} ((\Diamond_{[0.0, 20.0]}(y_5 - y_4 < 90.0) )  \vee   (\Diamond_{[0.0, 20.0]}(y_2 - y_1 > 15.0)   )\\
\text{CC15}:\; & \Box_{[0.0, 65.0]}\Diamond_{[0.0, 30.0]}\Box_{[0.0, 20.0]} ( y_5 - y_4 > 8.0 )\\
\text{CC16}:\; & \Box_{[0.0, 65.0]}\Diamond_{[0.0, 30.0]}\Box_{[0.0, 20.0]} ( y_4 - y_3 > 8.5 )\\
\text{CC17}:\; & \wedge_{i = 1, \ldots, 4}  \Box_{[0.0, 40.0]} ( y_{i+1} - y_i > 7.5 )\\
\text{CC18}:\; & \Box_{[0.0, 72.0]}\Diamond_{[0.0, 8.0]} ( \Box_{[0.0, 5.0]} ( y_2 - y_1 > 9.0 ) \Rightarrow \Box_{[5.0, 20.0]} ( y_5 - y_4 > 7.0 )) \\
\end{align*}
}

\subsection{Results}

\subsubsection{Parameters}

We first investigate the influence of key parameters of OSE (Algorithm~3): dimensions in which is the exploration procedure performed and feature levels. We consider the feature levels 1 to 3. We also consider various selections of dimensions $I$ for the AT and CC benchmarks. We investigated all possible combinations for the AT benchmark and we selected a few for the CC benchmark due to the large number of possible combinations. We set a maximum number of 20000 model simulations for the SI benchmark and 10000 model simulations for the AT and CC benchmarks. The overall success rate results can be seen in Table 1.

\begin{table}[h!]
\centering
\begin{tabular}{c | r r r } 
SI & level 1  & level 2 & level 3\\
\hline
$y$  &  128 &  139  & 129 \\
\hline
AT & level 1  & level 2 & level 3\\
\hline
speed & 376 & 384 & 380 \\
RPM & 341 &  348 & 347 \\
gear & 339 & 352 & 346\\
speed, RPM  & 373 & 378 & 349 \\
speed, gear & 385	& 378	& 363 \\
RPM, gear  & 382 & 377 & 347\\
all & 381 & 373 & 356 \\
\hline
CC & level 1  & level 2 & level 3\\
\hline
$y_1$ & 169 & 177 & 159 \\
$y_2$ & 163 & 172 & 157 \\
$y_3$ & 163 & 170 & 163 \\
$y_4$ & 162 & 169 & 163 \\
$y_5$ & 160 & 168 & 165 \\
$y_1, y_2$ & 164 & 152 & 130 \\
$y_4, y_5$ & 166 & 155 & 145 \\
$y_1, y_5$ & 172 & 158 & 147 \\
$y_1, y_2, y_3$ & 152 & 147 & 127 \\
$y_1, y_3, y_5$ & 155 & 151 & 138 \\
all & 152 & 136 & 131 \\
\hline
\end{tabular}
\caption{Parameters of OSE (Algorithm 3): overall success rates for various levels of features and dimensions in which is the exploration performed }
\end{table}


According to the results, the overall success rate can vary quite significantly.  The first major observation, that is consistent for all three benchmarks, is that the simplest feature selection---one level in one dimension---is not best performing strategy overall. We already observed the reason in Section 3.2: the shape of falsifying output signals may be to complex to be effectively searched for by a single-valued feature. For example, the selection of feature space  $\mathcal{F}^{y}_1$ for the SI benchmark resulted in the runs in which a counterexample to specification SI10  $(\Diamond_{[0, 30]} ( y > 6)) \Rightarrow (\Box_{[0, 30]} ( y>-6\;))$ was not generated at all, whereas for $\mathcal{F}^{y}_2$ and $\mathcal{F}^{y}_3$ were. A noticeable difference can also be observed  for specifications SI11--SI12, which include these "v-shaped" signals. However, features such as $\mathcal{F}^{y}_1$ and $\mathcal{F}^{\mathrm{speed}}_1$ are significantly more consistent in solving extreme-valued specifications, e.g., SI1--SI4 and AT1--AT2, in which multidimensional features such as $\mathcal{F}^{y}_3$ and $\mathcal{F}^{\mathrm{speed,gear}}_3$ struggled. Overall, we conclude that these simplest features should still be included in the exploration algorithm overall.

In the case of multidimensional outputs, we observe that lack of exploration along a feature of one level and one dimension can be  alleviated by considering multiple dimensions. For example, the success rate of exploration using features from $\mathcal{F}^{\mathrm{speed}}_2$ is very similar to the success rate using features from $\mathcal{F}^{\mathrm{speed,gear}}_1$ and $\mathcal{F}^{\mathrm{RPM,gear}}_1$ However, overall dimensionality of the feature space should not be too high wrt to system simulations budget. We observe a noticeable drop in the overall success rate for feature spaces of dimensionality 5 and higher for both AT and CC benchmarks.

Lastly, we observe that the choice of the output space dimensions in which the exploration is performed definitely matters. For the AT benchmark especially, the exploration based on feature spaces  $\mathcal{F}^{\mathrm{RPM}}_1$ and  $\mathcal{F}^{\mathrm{gear}}_1$ was quite poor (the overall performance is comparable to the exploration without the Voronoi bias, see Section 5.2.2) and did not improve much by increasing the level to $\mathcal{F}^{\mathrm{RPM}}_2$ and  $\mathcal{F}^{\mathrm{gear}}_2$. In comparison, the exploration along just one dimension of the output $\mathcal{F}^{\mathrm{speed}}_1$ and $\mathcal{F}^{\mathrm{speed}}_2$ were one of the best performing strategies. Interestingly enough, a combination of both $\mathcal{F}^{\mathrm{RPM},\mathrm{gear}}_1$ and $\mathcal{F}^{\mathrm{RPM},\mathrm{gear}}_2$ performed also very well.

\begin{table*}
\centering
{\scriptsize
\begin{tabular}{c | r r r r| r r r r | r r r r| r r r r | r r r r | r r r r} 

\multicolumn{25}{c}{\small{\textbf{Simple Integrator}}}\\
\hline
 & \multicolumn{4}{c|}{\textbf{UR}} &  \multicolumn{4}{c|}{\textbf{NR}} & \multicolumn{4}{c|}{\textbf{RW}} & \multicolumn{4}{c|}{\textbf{RG}} & \multicolumn{4}{c|}{$\mathbf{RRT}$} & \multicolumn{4}{c}{\textbf{OSE}}\\
\hline
1--4   &   0 &  0 &  0 & 0    & 10 & 10 & 10 & 10      &  0 &  0 &  3 &  3     &  0 &  0 &  1 &  1      & 10 & 10 & 10 & 10     & 10 &  9 & 10 & 10 \\
5--8   &  10 & 10 & 10 & 0    & 10 & 10 & 10 & 10      & 10 &  9 & 10 &  0     & 10 &  9 & 10 &  1      &  0 &  0 &  0 &  2     &  3 & 10 &  8 & 10 \\
9--12  &   0 &  0 &  0 & 4    & 10 & 10 & 10 &  3      &  2 &  3 &  6 & 10     &  0 &  3 &  5 & 10      &  2 &  0 &  0 &  0     & 10 & 10 & 10 &  9 \\
13--16 &   0 &  0 &  0 & 0    & 10 & 10 &  4 & 10      &  5 &  3 &  2 &  5     &  2 &  1 &  2 &  1      &  0 &  0 &  0 &  0     &  8 &  6 & 10 & 10 \\
\hline
 $\sum$ & \multicolumn{4}{c|}{\small 34} &  \multicolumn{4}{c|}{\small 147} & \multicolumn{4}{c|}{\small 71} & \multicolumn{4}{c|}{\small 56} & \multicolumn{4}{c|}{\small 44} & \multicolumn{4}{c}{\small 143}\\
\hline
\multicolumn{25}{c}{\small{\textbf{Automatic Transmission}}}\\
\hline
 & \multicolumn{4}{c|}{\textbf{UR}} &  \multicolumn{4}{c|}{\textbf{NR}} & \multicolumn{4}{c|}{\textbf{RW}} & \multicolumn{4}{c|}{\textbf{RG}} & \multicolumn{4}{c|}{$\mathbf{RRT}$} & \multicolumn{4}{c}{\textbf{OSE}}\\
\hline
1--4   &  0 &  0 &  0 &  5    & 10 & 10 & 10 & 10      &  3 &  5 & 10 &  9     &  0 &  5 & 10 & 10      &  0 &  0 &  0 &  2     & 10 & 10 & 10 & 10 \\
5--8   & 10 & 10 & 10 & 10    & 10 & 10 & 10 & 10      &  9 & 10 & 10 & 10     &  9 & 10 & 10 & 10      &  9 &  2 &  4 &  7     & 10 & 10 & 10 & 10 \\
9--12  &  0 &  1 & 10 &  0    & 10 & 10 & 10 & 10      &  9 &  7 & 10 &  2     &  8 &  7 & 10 &  0      &  0 &  0 & 10 &  0     & 10 & 10 & 10 & 10 \\
13--16 & 10 &  1 & 10 & 10    & 10 & 10 & 10 & 10      & 10 & 10 & 10 & 10     & 10 & 10 & 10 & 10      & 10 &  0 & 10 & 10     & 10 & 10 & 10 & 10 \\
17--20 & 10 & 10 &  0 & 10    & 10 & 10 & 10 & 10      & 10 & 10 & 10 & 10     & 10 & 10 & 10 & 10      & 10 & 10 &  9 & 10     & 10 & 10 & 10 & 10 \\
21--24 & 10 & 10 &  1 & 10    & 10 & 10 & 10 & 10      & 10 & 10 & 10 & 10     & 10 & 10 & 10 & 10      & 10 & 10 &  5 & 10     & 10 & 10 & 10 & 10 \\
25--28 & 10 & 10 & 10 & 10    & 10 & 10 & 10 & 10      & 10 &  9 &  9 &  5     & 10 &  9 & 10 &  2      &  8 &  2 &  0 &  0     & 10 & 10 & 10 &  3 \\
29--32 & 10 & 10 &  0 &  0    & 10 & 10 & 10 &  0      & 10 & 10 &  7 &  5     & 10 & 10 &  6 &  4      &  6 & 10 &  9 &  0     & 10 & 10 & 10 &  6 \\
33--36 &  1 &  0 &  0 & 10    &  5 &  5 &  5 & 10      &  9 & 10 & 10 & 10     &  7 &  7 &  7 & 10      &  3 &  2 &  2 &  7     & 10 &  9 &  9 & 10 \\
37--40 & 10 &  0 &  2 &  6    &  9 & 10 &  2 &  3      & 10 &  2 &  3 & 10     & 10 &  1 &  2 &  9      &  2 &  0 &  0 &  1     & 10 & 10 &  8 & 10 \\
\hline
 $\sum$ & \multicolumn{4}{c|}{\small 237} &  \multicolumn{4}{c|}{\small 359} & \multicolumn{4}{c|}{\small 346} & \multicolumn{4}{c|}{\small 323} & \multicolumn{4}{c|}{\small 188} & \multicolumn{4}{c}{\small 385}\\
\hline
\multicolumn{25}{c}{\small{\textbf{Chasing Cars}}}\\
\hline
 & \multicolumn{4}{c|}{\textbf{UR}} &  \multicolumn{4}{c|}{\textbf{NR}} & \multicolumn{4}{c|}{\textbf{RW}} & \multicolumn{4}{c|}{\textbf{RG}} & \multicolumn{4}{c|}{$\mathbf{RRT}$} & \multicolumn{4}{c}{\textbf{OSE}}\\
\hline
1--4   &  0 &  0 &  0 &  0    & 10 & 10 & 10 & 10      &  1 &  0 &  0 &  0     &  1 &  0 &  0 &  0      &  0 &  0 &  0 &  0     & 10 & 10 & 10 &  9 \\
5--8   & 10 & 10 &  7 &  1    & 10 & 10 & 10 & 10      & 10 & 10 & 10 & 10     & 10 & 10 & 10 & 10      &  4 &  2 &  1 &  0     & 10 & 10 & 10 & 10 \\
9--12  &  0 &  0 &  0 &  0    & 10 & 10 & 10 & 10      & 10 &  7 &  8 &  8     & 10 &  8 &  7 &  7      &  1 &  0 &  0 &  0     & 10 &  8 & 10 &  8 \\
13--16 &  0 &  0 &  0 &  0    & 10 & 10 & 10 &  9      & 10 &  8 &  5 &  0     & 10 &  7 &  6 &  1      &  0 &  0 &  0 &  0     & 10 & 10 & 10 & 10 \\
16--18 &  8 &  0 &    &       & 10 & 10 &    &         & 10 &  9 &    &        & 10 & 10 &    &         &  0 &  0 &    &        & 10 & 10 &    &   \\
\hline
 $\sum$ & \multicolumn{4}{c|}{\small 36} &  \multicolumn{4}{c|}{\small 179} & \multicolumn{4}{c|}{\small 116} & \multicolumn{4}{c|}{\small 117} & \multicolumn{4}{c|}{\small 8} & \multicolumn{4}{c}{\small 175}\\
\end{tabular}
}
\caption{Exploration: Success rate for uniform random sampling (UR), nonuniform random sampling (NR), random walk (RW), random graph (RG), rapidly exploring random trees (RRT), and Output Space Exploration (OSE). $\sum$ denotes a sum of successful falsifications over all specifications after a maximum number of simulations.}
\end{table*}

\subsubsection{Exploration}
\label{sec:exploration}

Next, we compare the performance of the following exploration-based algorithms. 
\paragraph{Uniform random (UR)} The algorithm tests input signals that are sampled uniformly randomly from the set of input values $U$.
\paragraph{Nonuniform random (NR)} The algorithm tests input signals that are sampled randomly from a set inspired by FalStar~\cite{falstar}. Assume a rectangular set of input values $U = [u_\mathrm{min},u_\mathrm{max}]$. The algorithm generates the input signals incrementally, i.e., $u = u^1u^2 \ldots .$ Values of the $i$-th dimension of the segment $u^j$ at a given \emph{level} $l \in \{0, 1 , \ldots, L_\mathrm{max}\}$ are given as 
\begin{equation}
u_i^j = u_{\mathrm{min},i} + p_l(u_{\mathrm{max},i} - u_{\mathrm{min},i}),
\end{equation}
where
\begin{equation}
p_l \in \left\{   \frac{2^j+1}{2^l} \mid j = 0,1, \ldots \wedge j \leq \frac{2^l-1}{2} \right\},
\end{equation}
The total level of the segment $u^j$ is the sum of levels in each dimension of $u.$  The probability of selecting a particular value $u^j = (u_1^j, \ldots, u_n^j)$ is    $c 2^{-L}$, where $L$ is the total level of the segment $u^j$ and where $c$ is the normalization constant computed over all possible values of $u^j$ up to the maximal considered total level $L_\mathrm{max}$. We picked  $L_\mathrm{max} = 10$ for the SI benchmark, $L_\mathrm{max} = 6$ for the AT benchmark, and  $L_\mathrm{max} = 5$ for the CC benchmark. The number of control nodes  for an input signal of a given level (i.e., its length) were selected as $[15, 15, 10, 8, 6, 5, 5, 4, 4, 3, 2]$ for the SI benchmark, $[3, 3, 2, 2, 1, 1, 1]$ for the AT benchmark, and $[10, 10, 8, 8, 4, 4]$ for the CC benchmark. 
i.e., $p_0 \in \{0,1\}, p_1 \in \{\frac{1}{2}\}, p_3 \in \{\frac{1}{4},\frac{3}{4}\} \ldots .$ 

\paragraph{Random walk (RW)} The algorithm selects the first input signal uniformly randomly and subsequently repeatedly modifies the input signal using the \emph{input selection} rule of Algorithm 3 (OSE).
\paragraph{Random walk (RG)} The algorithm selects the first input signal uniformly randomly. In each subsequent iteration, the algorithm selects uniformly randomly one input signal from the set of
all output signals that were previously generated during the run of the algorithm. Then it modifies the selected input signal using the \emph{input selection} rule of Algorithm 3 (OSE).
\paragraph{Rapidly Exploring Random Tree Algorithm (RRT)} The algorithm follows Algorithm 1. A random target $y'$ from $Y$ at a time instant $t$ is generated using~\eqref{est}. The time instant $t$ itself is selected uniformly randomly from the set $\{0, h, 2h, \ldots \},$ where $h$ is a time step between discontinuities of the input. 

For a given target $(y',t')$, a selected node $(y,t)$ minimizes
\begin{equation}
\mu(y,y') = \sum_{i \in m} \left(\frac{y_i - y'_i}{\mathrm{UB}_i(t) - \mathrm{LB}_i(t)}\right)^2
\end{equation}
among all of the nodes of the tree at the time instant $t$. Inputs are generated uniformly randomly from $U$.

\begin{figure*}[h!]
\centering
\begin{subfigure}{0.32\textwidth}
\includegraphics[width=\textwidth]{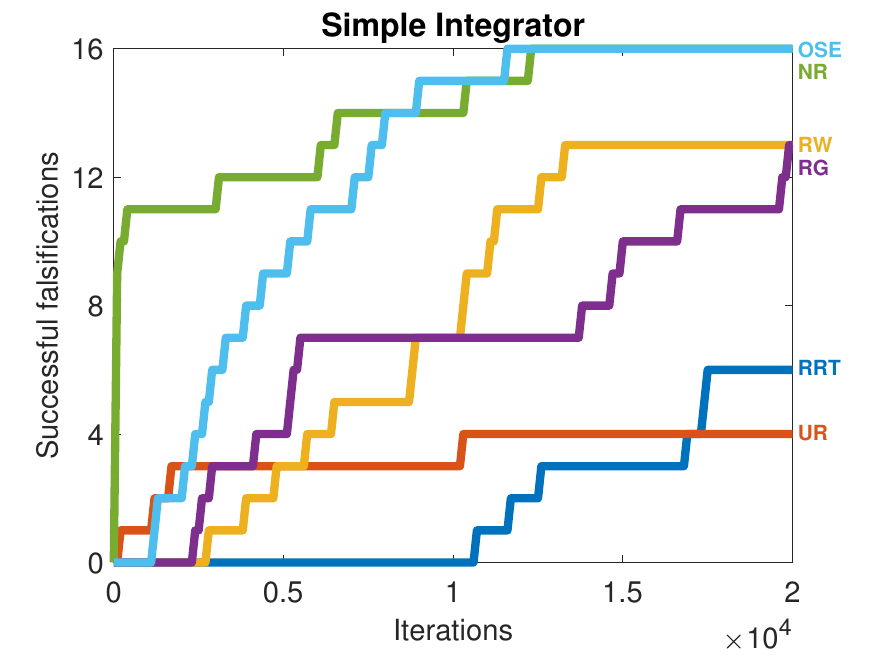}
\end{subfigure}
\begin{subfigure}{0.32\textwidth}
\includegraphics[width=\textwidth]{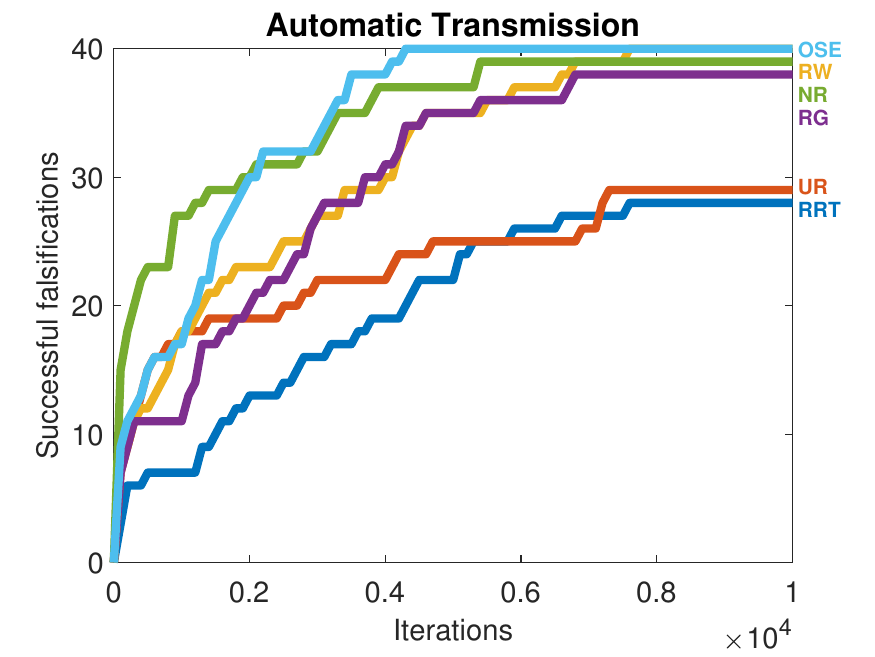}
\end{subfigure}
\begin{subfigure}{0.32\textwidth}
\includegraphics[width=\textwidth]{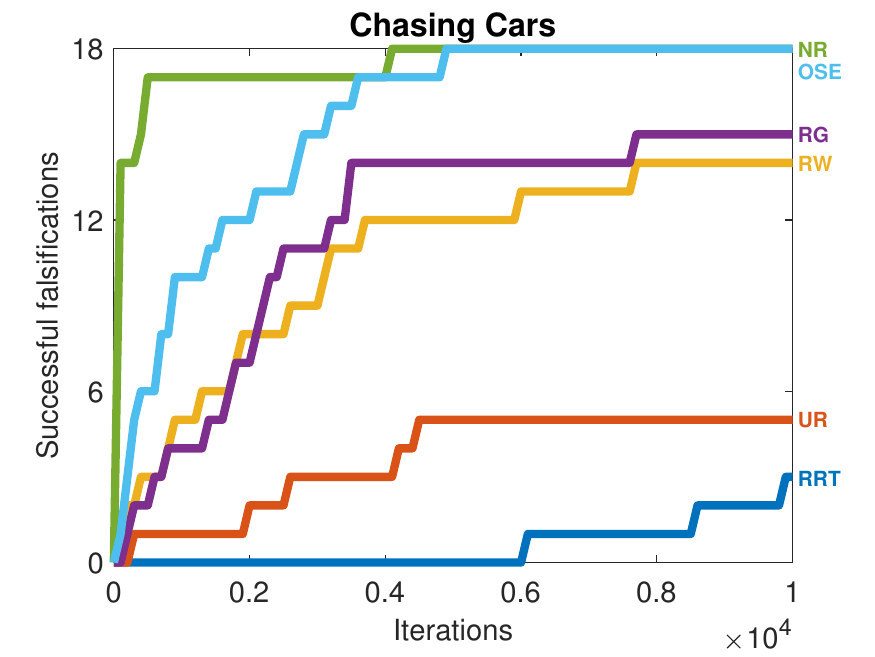}
\end{subfigure}
\caption{Convergence of exploration algorithms: uniform random (UR, red), nonuniform random (NR, green), random walk (RW, yellow), random graph (RG, purple) , Algorithm 1 (RRT, dark blue),  Output Space Exploration (OSE, light blue). }
\end{figure*}

\paragraph{Output Signal Exploration (OSE)} The algorithm follows Algorithm 3. The parameters of OSE were selected as described in the implementation section, i.e., the selection of feature levels was randomized according to probabilities $p_1 = \frac{4}{7}, p_2 = \frac{2}{7}$, and $p_3 = \frac{1}{7}$. We did not randomized the selection of dimensions $I$ and kept it fixed. We selected dimensions $I$ based on the computational experiments from the previous subsection: the exploration was restricted to the first output for all benchmarks, i.e., speed for the AT benchmark and $y_1$ for the CC benchmark.

\smallskip
We allowed 20000 model simulations for the SI benchmark and 10000 model simulations for the AT and CC benchmarks for algorithms UR, NR, RG, RW, and OSE (we counted only the unique inputs for the algorithm NR and discarded all repeats). For the RRT algorithm, we considered 40000 model simulations for the SI benchmark and 20000 model simulations for the AT and CC. This increase reflects the fact that the partial simulations are in average half as long, see Section 3.2. All runs of all algorithms  were repeated 10 times each. We evaluated the robustness of resulting output signals wrt specifications SI1-SI16, AT1-AT40, and CC1-CC18 to demonstrate how varied the output signals are by finding falsifying output signals. Let us compare the results of all algorithms, see Table 2 and Figure 4.

\paragraph*{Uniform (UR) and nonuniform (NR) random sampling} Concerning the UR algorithm, it is, according to the benchmarks, one of the least-performing strategies for falsification. It seems that the space of input signals is simply too large to make the uniform random sampling effective strategy. In comparison, the NR algorithm is actually one of the best-performing strategies. Its assumption that falsifying input signals are often simple in shape holds for many benchmarks, as was already observed in~\cite{falstar}. This makes NR extremely efficient compared to other investigated exploration strategies, often requiring a small fraction of simulations to detect a counterexample. However, NR has an inherent chance of failing to detect some counterexamples because they can require an input signal that is rarely generated (or cannot be generated) for the chosen discretization. An illustrative example of such a case can be observed here by benchmarks SI12, SI15,  AT33--AT35, AT38--AT40, and especially AT32, for which no falsifying counterexamples were found by the NR algorithm.

\paragraph*{Rapidly exploring random tree (RRT)} The RRT algorithm is the least-performing algorithm overall. Its small incremental steps, in combination with a relatively small number of  model simulations, is certainly detrimental to its performance. We selected the input selection in this experiment as it is since it is representative of the ones used for falsification in~\cite{branicky2006sampling,dreossi2015efficient} and the ones used for motion planning in general. But the context of motion planning, in which the simulations are often relatively cheap and input steps must be taken small due to the presence of obstacles, proves to be very different from falsification. It should be noted that these issues can be addressed by allowing either significantly a larger number of simulations or by allowing larger input signal variations in a single step. Note that the second change brings the RRT algorithm in its implementation closer how the algorithm OSE works. 

\paragraph*{Algorithms 3 and its simplifications (RG, RW)} The RG/RW algorithms are variants of OSE with a selection step that is not based on the Voronoi bias. In this comparison, we observe a benefit of the Voronoi bias: OSE is consistently able to detect counterexamples to specifications such as SI1--SI4, SI8--SI10, AT1--AT2, AT12, AT28, AT38--AT39 and CC1--CC4, CC15--CC1, whereas the RW and RG algorithms are not. The overall convergence rate of OSE is also noticeably faster. 

In comparison to NR, OSE and NR have similar overall success rates: NR struggled with some specifications, namely CC12, CC15, AT32--35 and AT39--40, OSE with specifications SI5 and AT28. The NR algorithm was significantly faster in detecting some counterexamples early on due to exploiting a typical shape of falsifying input signals, although OSE seems to eventually always catch on, and thus, the overall convergence rate is similar as well.

\smallskip

Overall, no strategy is entirely superior to others, although concerning all strategies compared here,  NR and OSE performed significantly better than the rest. OSE is noticeably more successful in finding counterexamples for the AT model, since NR struggled with some AT specifications, but NR is a bit more consistent in finding counterexamples to CC specifications. In addition, NR and OSE seem to complement each other well. The NR algorithm can be used to cheaply map the space of output signals using a simple input space grid and find counterexamples tied to such inputs. Subsequently, OSE can further refine the generated output signals and attempt to detect the remaining specifications that are difficult to find for NR, e.g., SI12, SI15,  AT32--AT35, and AT39--AT40 in our particular experiments.

\begin{table*}
\centering
{\tiny
\begin{tabular}{c | r r r r| r r r r | r r r r| r r r r | r r r r | r r r r} 
\small Ex & \multicolumn{8}{c|}{\small\textbf{SHC}} &  \multicolumn{8}{c|}{\small\textbf{CMA-ES}} & \multicolumn{8}{c}{\small\textbf{DE}} \\
\hline
 & \multicolumn{4}{c|}{\small\textbf{Succ}} &  \multicolumn{4}{c|}{\small\textbf{Eval}} & \multicolumn{4}{c|}{\small\textbf{Succ}} & \multicolumn{4}{c|}{\small\textbf{Eval}} & \multicolumn{4}{c|}{\small \textbf{Succ}} & \multicolumn{4}{c}{\small \textbf{Eval}} \\
\hline
\multicolumn{25}{c}{\small{\textbf{Simple Integrator}}}\\
\hline
1--4  &   10 & 10 & 10 & 10    & 247 & 260 &  71 &  70      & 10 & 10 & 10 & 10     & 257 & 283 & 148 & 143      & 10 & 10 & 10 & 10    & 730 & 664 & 379 & 358 \\ 
5--8  &   10 & 10 & 10 & 10    & 219 & 211 & 208 & 224      & 10 & 10 & 10 & 10     & 160 & 111 & 149 & 183      & 10 & 10 & 10 & 10    & 110 & 142 & 168 & 382 \\
9--12  &  10 & 10 & 10 & 10    & 138 & 155 & 134 &  93      & 10 & 10 & 10 &  8     & 199 & 335 & 211 & 651      & 10 & 10 & 10 & 10    & 383 & 464 & 316 & 597 \\
13--16 &  10 & 10 & 10 & 10    & 379 & 127 & 175 & 183      &  9 & 10 & 10 & 10     & 567 & 338 & 167 & 212      & 10 & 10 & 10 & 10    & 751 & 350 & 406 & 372 \\
\hline
 $\sum$ & \multicolumn{4}{c|}{\small 160} &  \multicolumn{4}{c|}{\small 2894} & \multicolumn{4}{c|}{\small 157} & \multicolumn{4}{c|}{\small 4113} & \multicolumn{4}{c|}{\small 158} & \multicolumn{4}{c}{\small 6572}\\
\hline
\multicolumn{25}{c}{\small{\textbf{Automatic Transmission}}}\\
\hline
1--4  &   10 & 10 & 10 & 10    & 102 &  73 &  62 &  74      & 10 & 10 & 10 & 10     & 483 & 336 & 247 & 208       &  9 & 10 & 10 & 10    & 535 & 170 &   74 & 120 \\ 
5--8  &   10 & 10 & 10 & 10    & 253 & 205 & 186 & 123      & 10 & 10 & 10 & 10     & 235 & 156 &  66 & 107       & 10 & 10 & 10 & 10    & 174 &  79 &  148 & 113 \\
9--12  &  10 & 10 & 10 & 10    & 121 &  89 &  31 &  67      & 10 & 10 & 10 & 10     & 298 & 295 &  73 & 436       &  9 & 10 & 10 & 10    & 463 & 179 &  105 & 389 \\
13--16 &  10 & 10 & 10 & 10    &  38 & 437 &  15 &  55      & 10 & 10 & 10 & 10     & 212 & 386 &  11 &  72       &  7 &  9 & 10 & 10    & 884 & 802 &   30 &  53 \\
17--20 &  10 & 10 & 10 & 10    &   4 & 153 &  55 &  94      & 10 & 10 &  8 & 10     &   2 &  65 & 688 & 195       & 10 & 10 & 10 & 10    &  13 &  49 &  599 & 296 \\
21--24 &  10 & 10 & 10 & 10    & 189 &   7 & 490 & 235      & 10 & 10 &  5 & 10     & 103 &   4 &1551 & 126       & 10 & 10 &  7 & 10    & 116 &  12 & 1578 & 206 \\
25--28 &  10 & 10 & 10 & 10    &  82 & 159 &  71 & 150      & 10 & 10 & 10 & 10     &  78 & 177 & 117 & 291       & 10 &  9 & 10 & 10    &  95 & 507 & 190  & 829 \\
29--32 &  10 & 10 & 10 &  8    &  56 &  23 & 163 & 965      & 10 & 10 & 10 &  6     &  53 &  62 & 395 &1489       & 10 & 10 & 10 &  4    & 102 & 309 & 388  &1909 \\
33--36 &  10 &  8 &  6 & 10    & 274 &1030 &1588 & 778      &  9 &  6 &  0 & 10     & 679 &1412 &2500 & 134       &  3 &  0 &  2 & 10    &1794 &2500 &2129  & 313 \\
37--40 &  10 &  6 &  4 & 10    & 805 &1331 &1826 & 234      & 10 &  9 &  4 & 10     & 354 &2359 &2141 & 354       & 10 &  2 &  4 & 10    & 194 &2219 &1961  & 503 \\
\hline
 $\sum$ & \multicolumn{4}{c|}{\small 382} &  \multicolumn{4}{c|}{\small 12687} & \multicolumn{4}{c|}{\small 367} & \multicolumn{4}{c|}{\small 19096} & \multicolumn{4}{c|}{\small 345} & \multicolumn{4}{c}{\small 23126}\\
\hline
\multicolumn{25}{c}{\small{\textbf{Chasing Cars}}}\\
\hline
1--4  &   10 & 10 & 10 & 10    & 127 & 152 & 222 & 949      & 10 & 10 &  7 &  0     & 714 & 891 &1432 &2500       & 10 & 10 & 10 &  5    & 284 & 275 & 365 &1570 \\ 
5--8  &   10 &  0 &  1 &  7    &  72 &2500 &2259 & 932      &  7 &  0 &  0 &  3     & 786 &2500 &2500 &1848       &  2 &  0 &  1 &  0    &2006 &2500 &2252 &2500 \\
9--12  &  10 &  1 &  0 &  0    &  52 &2294 &2500 &2500      & 10 &  0 &  0 &  0     & 287 &2500 &2500 &2500       & 10 &  0 &  0 &  0    & 110 &2500 &2500 &2500 \\
13--16 &  10 & 10 & 10 & 10    &  75 &  71 & 176 & 465      & 10 & 10 &  9 &  3     & 387 & 446 & 845 &2087       & 10 & 10 & 10 & 10    & 140 & 199 & 435 & 489 \\
17--20 &  10 & 10 &    &       &  94 & 197 &     &          & 10 & 10 &    &        & 260 & 613 &     &           & 10 & 10 &    &       & 118 & 218 &     &     \\
\hline
 $\sum$ & \multicolumn{4}{c|}{\small 129} &  \multicolumn{4}{c|}{\small 15634} & \multicolumn{4}{c|}{\small 98} & \multicolumn{4}{c|}{\small 25597} & \multicolumn{4}{c|}{\small 108} & \multicolumn{4}{c}{\small 20899}\\
\hline
\end{tabular}
}
\caption{Optimization methods: Success rate (Succ) in 10 runs and mean number of simulations (Eval) over 10 runs for stochastic hill climbing (SHC), differential evolution (DE) and CMA-ES. $\sum$ denotes a sum of successful falsifications and a sum of a mean number of simulations over all specifications.}
\end{table*}

\subsubsection{Optimization}
\label{sec:optimization}

In the last part of the computational experiments, we compare the exploration-based algorithms from the previous experiment with optimization-based methods. Pure exploration-based algorithms do not depend on an STL specification under investigation. Thus, they can test multiple STL specifications in a single run. To demonstrate the benefits of this approach, we now run optimization-based methods for each STL specification separately. We consider the following three optimization methods:

\paragraph{Stochastic hill climbing (SHC)} The algorithm starts optimization from the input signal that is sampled uniformly randomly from the specified set of inputs. Then it modifies the input in each step and accepts this step provided that the robustness is lesser or equal to the robustness value of the original input signal. In our implementation, the modification of input signals is based on the same generation of disturbances as in OSE, i.e., its generation of inputs is based on the Cauchy distribution with the same scaling and the same crossover probability.

\paragraph{Differential evolution (DE)} Differential evolution is a family of population-based global optimization methods that are inspired by the Nelder-Mead algorithm~\cite{de}. We selected the variant DE/best/1/bin with a population $N = 50$, a scale parameter $F = 0.5$, and a crossover probability $CR = 0.5.$ See~\cite{de} for more details. 

\paragraph{Covariance matrix adaptation evolution strategy (CMA-ES)} CMA-ES is an evolutionary strategy primarily designed for a robust local optimization~\cite{cmaes}. However, it has been successfully applied for global optimization problems as well. CMA-ES is based on a self-adaptation of the covariance matrix of the normal distribution from which the population members are generated, which steers the population toward the local optimum.

\smallskip
We used all three algorithms for each benchmark. For each specification, we ran the optimization algorithms from a random point/population generated from the uniform distribution over the set of allowed inputs. We allowed 2500 model simulations for falsification of each specification. Note that this is in contrast to the experiments in Section~\ref{sec:exploration}, where the number of model simulations is shared across all requirements for a given benchmark model. We repeated each falsification 10 times. The results are in Table~3.

While specifications for the SI benchmark are quite simple for optimization-based falsification, the AT and CC benchmarks included specifications that are difficult to optimize for all three optimization methods according to the results. Thus, none of these optimization-based algorithms actually outperformed the overall success rates of the exploration-driven algorithms: nonuniform random sampling (NR) and Output Space Exploration (OSE). This strongly indicates that a generally successful optimization tool must include both optimization and exploration components.

\section{Related Work}

One of the most prominent and successful approach in the state-of-the-art software tools for the falsification~\cite{arch} of STL requirements of black-box models uses optimization~\cite{breach,psystaliro,athena,foresee}, which detects design errors by finding a global minimum of robustness~\cite{stl} (or other similar guidance metric associated with a given error STL specification that is being tested such as fitness functions~\cite{athena} and QB-robustness~\cite{foresee}). 


Numerous global and local optimization techniques have been used for falsification, such as gradient descent~\cite{grad1,grad2}, Nelder-Mead~\cite{breach}, simulated annealing~\cite{nghiem2010monte,aerts}, CMA-ES~\cite{breach,falstar}, Bayesian optimization~\cite{mathesen2019falsification,psystaliro}, ant-colony optimization~\cite{ants}, cross-entropy method~\cite{crossentropy}, and others.
Approaches that are tuned towards falsifying multiple requirements at the same time are~\cite{peltomaki2024testing,liden2022multi}. This capability is inherently given in our approach already.

Coverage metrics haven been explored in the context of falsification,
for example by looking at all (discrete) combinations of modes of hybrid systems~\cite{dokhanchi2015requirements}
or by hyperplane cutting~\cite{adimoolam2017classification}.
Other methods probabilistically search inputs of the system to identify counterexamples. Software tool~\cite{falstar} uses Monte Carlo tree search (MCTS) that exploits the fact that inputs producing the counterexample behavior tend to be simple more often than not, e.g., inputs that consist of extreme values. Hence, the tool generates these inputs with a greater probability.

Another family of method constructs surrogate models using the outputs of a black-box model. Then, they aim to find falsifying counterexamples for the surrogate models and refine these surrogate models provided that the produced counterexamples are spurious for the original black-box model. The proposed surrogate models includes surrogate models based on system identification~\cite{SI}, Mealy machines~\cite{mealy}, neural networks~\cite{nn}, and Koopman models~\cite{koopman}.

\section{Conclusion}
We proposed a novel falsification algorithm based on the exploration in the space of output signals driven by the Voronoi bias. In comparison to the approaches proposed in the past, our algorithm does not employ exploration based on coverage inputs nor uses robustness to steer the exploration towards the falsifying behavior. As we demonstrated in our computational experiments, this allows our algorithm to handle falsification of specifications that are difficult to find using other previously suggested falsification methods while still achieving a good overall performance among all compared methods.


\bibliographystyle{plain}
\bibliography{ref_ar}

\end{document}